\documentclass[aps,preprint,showpacs]{revtex4}

\begin{document}
%\draft

\title{
Quantum discord for two-qubit CS  states:
Analytical solution 
}

\author{
M~A~Yurishchev
%\thanks{Electronic address: yur@itp.ac.ru} %%
}
\email{yur@itp.ac.ru} %%
\affiliation{ %%
%\address{ %%
Institute of Problems of Chemical Physics of the Russian Academy of Sciences,
%School blvd 1B-52,
142432 Chernogolovka,
Moscow Region,
Russia 
}

%\date{\today}

%\maketitle %%

\begin{abstract}
We found an universal {\em local\/} orthogonal transformation
which transforms any centrosymmetric (CS) density matrix $\rho_{CS}$
into the X density matrix $\rho_X$:
$H\otimes H\rho_{CS}H\otimes H=\rho_X$, where $H$ is the Hadamard
transformation.
Since quantum discord is invariant under the local unitary
transformations, this remarkable property allows to get the discord
of general two-qubit CS states in an analytical form using the
corresponding well-known formulas for the X states. 
Examples of systems with the CS density matrices are given, including
XXZ spin model with the Dzyaloshinsky-Moriya interaction,
a gas of spin-carrying particles in closed nanopore,
and a family of pseudopure states.
\end{abstract}

%\medskip %%
%PACS: 03.65.Ud, 03.67.-a, 75.10.Jm\\ %%

\maketitle %%

%======================================================================

{\em Introduction.\/} Quantum discord, $Q$, is a more general measure
of quantum correlation than entanglement.
Discord is important in quantum information theory, quantum
metrology, condensed matter physics, and so on \cite{MBCPV12}.
The evaluation of discord is a very hard problem.
Even for two-qubit systems, the analytical calculations are restricted
to the so-called X states.
The term ``X states'' has been introduced in 2007 \cite{YE07} and denotes
the $4\times4$ density matrices which may have non-zero entries only
along the main diagonal and anti-diagonal.
Algebraic properties of such matrices were studied by Rau \cite{R09}.
Analytical solution for the quantum discord of X density matrices
has been obtained in Ref.~\cite{ARA10}.
On the other hand, there is a number of physical systems for which the
density matrices have different forms, e.~g., centrosymmetric ones.
The $n\times n$ CS matrix is defined by the relations for its elements:
$a_{ij}=a_{n-i+1,n-j+1}$
(about the CS matrices see, for example, the reviews \cite{A98} and
references therein).
In this communication we establish a connection between CS and X matrices
through the local orthogonal transformation.
Thanks to the fact \cite{MBCPV12} that the discord (and the
entanglement) does not change its value under such transformations,
we as a result obtain a possibility for the analytic calculation of
discord in arbitrary two-qubit CS states.

{\em CS and X matrices.\/} In quantum mechanics the density matrix must
be Hermitian, non-negativity defined, and have unit trace.
A general $4\times4$ CS density matrix can be written as
\begin{equation}
   \label{eq:rho-CS}
   \rho_{CS}=\left(
      \begin{array}{cccc}
      p_1&p_2+ip_3&p_4+ip_5&p_6\\
      p_2-ip_3&{1\over2}-p_1&p_7&p_4-ip_5\\
      p_4-ip_5&p_7&{1\over2}-p_1&p_2-ip_3\\
      p_6&p_4+ip_5&p_2+ip_3&p_1
      \end{array}
   \right).
\end{equation}
It contains seven real parameters $p_1, \ldots , p_7$.
For the X state we have
\begin{equation}
   \label{eq:rho-X}
   \rho_X=\left(
      \begin{array}{cccc}
      \rho_{11}&0&0&\rho_{14}\\
      0&\rho_{22}&\rho_{23}&0\\
      0&\rho_{32}&\rho_{33}&0\\
      \rho_{41}&0&0&\rho_{44}
      \end{array}
   \right)
   =\left(
      \begin{array}{cccc}
      q_1&0&0&q_4+iq_5\\
      0&q_2&q_6+iq_7&0\\
      0&q_6-iq_7&q_3&0\\
      q_4-iq_5&0&0&1-q_1-q_2-q_3
      \end{array}
   \right).
\end{equation}
This matrix has also seven real parameters $q_1, \ldots , q_7$.
Let us consider the transformation
\begin{equation}
   \label{eq:R}
   R=H\otimes H
   ={1\over2}\left(
      \begin{array}{rrrr}
      1&1&1&1\\
      1&-1&1&-1\\
      1&1&-1&-1\\
      1&-1&-1&1
      \end{array}
   \right),
\end{equation}
where
\begin{equation}
   \label{eq:H}
   H={1\over\sqrt{2}}\left(
      \begin{array}{rr}
      1&1\\
      1&-1
      \end{array}
   \right)
\end{equation}
is the Hadamard transform.
The matrix $R$ is orthogonal and $R^T=R$ (the superscript $T$ denotes
a transposition).
Taking the matrices (\ref{eq:rho-CS})--(\ref{eq:R}) and performing
straightforward calculations we establish the following relations 
\begin{equation}
   \label{eq:CS-X}
   H\otimes H\rho_{CS}H\otimes H=\rho_X
\end{equation}
and
\begin{equation}
   \label{eq:X-CS}
   H\otimes H\rho_XH\otimes H=\rho_{CS} .
\end{equation}
Here the parameters of both matrices are related as
\begin{eqnarray}
   \label{eq:p-q}
   &&\rho_{11}=q_1={1\over4}+p_2+p_4+{1\over2}\,(p_6+p_7),
   \nonumber\\
   &&\rho_{22}=q_2={1\over4}-p_2+p_4-{1\over2}\,(p_6+p_7),
   \nonumber\\
   &&\rho_{33}=q_3={1\over4}+p_2-p_4-{1\over2}\,(p_6+p_7),
   \nonumber\\
   &&\rho_{44}=1-q_1-q_2-q_3={1\over4}-p_2-p_4+{1\over2}\,(p_6+p_7),
   \\
&&{\rm Re}\rho_{14}={\rm Re}\rho_{41}=q_4=-{1\over4}+p_1+{1\over2}\,(p_6-p_7),
   \nonumber\\
   &&{\rm Im}\rho_{14}=-{\rm Im}\rho_{41}=q_5=-p_3-p_5 ,
   \nonumber\\
&&{\rm Re}\rho_{23}={\rm Re}\rho_{32}=q_6=-{1\over4}+p_1-{1\over2}\,(p_6-p_7),
   \nonumber\\
   &&{\rm Im}\rho_{23}=-{\rm Im}\rho_{32}=q_7=p_3-p_5  
   \nonumber
\end{eqnarray}
and, vice versa,
\begin{eqnarray}
   \label{eq:q-p}
   &&p_1={1\over4}+{1\over2}\,(q_4+q_6),
   \nonumber\\
   &&p_2=-{1\over4}+{1\over2}\,(q_1+q_3),
   \nonumber\\
   &&p_3=-{1\over2}\,(q_5-q_7),
   \nonumber\\
   &&p_4=-{1\over4}+{1\over2}\,(q_1+q_2),
   \\
   &&p_5=-{1\over2}\,(q_5+q_7),
   \nonumber\\
   &&p_6={1\over4}-{1\over2}\,(q_2+q_3-q_4+q_6),
   \nonumber\\
   &&p_7={1\over4}-{1\over2}\,(q_2+q_3+q_4-q_6).
   \nonumber
\end{eqnarray}
It is easy to proof the same using the Bloch forms for the density
matrices $\rho_{CS}$ and $\rho_X$.
Indeed, expanding the density matrix (\ref{eq:rho-CS}) on the Pauli
matrices one obtains 
\begin{eqnarray}
   \label{eq:rhoCS-Bloch}
   \rho_{CS}&=&{1\over4}[1 + 4p_4\sigma_x\otimes1 + 4p_21\otimes\sigma_x
   + 2(p_6 + p_7)\sigma_x\otimes\sigma_x 
   + 2(p_7 - p_6)\sigma_y\otimes\sigma_y 
   \nonumber\\
   &+& (4p_1 - 1)\sigma_z\otimes\sigma_z 
   - 4p_3\sigma_z\otimes\sigma_y 
   - 4p_5\sigma_y\otimes\sigma_z] .
\end{eqnarray}
Performing the Hadamard transformations and taking into
account that $H\sigma_xH=\sigma_z$, $H\sigma_yH=-\sigma_y$, and
$H\sigma_zH=\sigma_x$, we obtain
\begin{eqnarray}
   \label{eq:HHrhoCSHH-Bloch}
   H\otimes H\rho_{CS}H\otimes H&=&{1\over4}[1 + 4p_4\sigma_z\otimes1
   + 4p_21\otimes\sigma_z
   + 2(p_6 + p_7)\sigma_z\otimes\sigma_z 
   + 2(p_7 - p_6)\sigma_y\otimes\sigma_y 
   \nonumber\\
   &+& (4p_1 - 1)\sigma_x\otimes\sigma_x 
   + 4p_3\sigma_x\otimes\sigma_y 
   + 4p_5\sigma_y\otimes\sigma_x]=\rho_X .
\end{eqnarray}
The last equality follows from the Bloch form for the matrix
(\ref{eq:rho-X})
\begin{eqnarray}
   \label{eq:rhoX-Bloch}
   \rho_X&=&{1\over4}\{ 1
   - [1 - 2(q_1 +q_2)]\sigma_z\otimes1
   - [1 - 2(q_1 +q_3)]1\otimes\sigma_z
   + 2(q_4 + q_6)\sigma_x\otimes\sigma_x 
   \nonumber\\
   &+& 2(q_6 - q_4)\sigma_y\otimes\sigma_y 
   + [1 - 2(q_2 + q_3)]\sigma_z\otimes\sigma_z 
   - 2(q_5 - q_7)\sigma_x\otimes\sigma_y 
   \nonumber\\
   &-& 2(q_5 + q_7)\sigma_y\otimes\sigma_x\}
\end{eqnarray}
and the relations (\ref{eq:p-q}).

As a result, we conclude that the quantum discord of the state
$\rho_{CS}$ is expressed through the discord of the state $\rho_X$:
\begin{equation}
   \label{eq:Q}
   Q(\rho_{CS})=Q(\rho_X),
\end{equation}
where the entries of $\rho_X$ are given by Eqs.~(\ref{eq:p-q}).
Together with the analytical formulas for the discord of X states
\cite{ARA10}, this completes our solution.

{\em Physical examples.\/} As a first illustration, consider the
anisotropic XXZ model with the Dzyaloshinsky-Moriya interaction.
When the Dzyaloshinsky vector ${\bf D}$ is oriented along $x$-direction,
the Hamiltonian for the two-qubit chain reads \cite{CZ10} 
\begin{equation}
   \label{eq:Hxxz-dm}
   {\cal H} = J\sigma_1^x\sigma_2^x 
   + J\sigma_1^y\sigma_2^y 
   + J_z\sigma_1^z\sigma_2^z 
   + D_x(\sigma_1^y\sigma_2^z-\sigma_1^z\sigma_2^y).
\end{equation}
Here $J$ and $J_z$ are the coupling constants and $\sigma_i^\alpha$
($i=1,2$ and $\alpha=x,y,z$) are the Pauli matrices.  
In open form the Hamiltonian is given as
\begin{equation}
   \label{eq:Hxxz-dm1}
   {\cal H}=\left(
      \begin{array}{cccc}
      J_z&iD_x&-iD_x&0\\
      -iD_x&-J_z&2J&iD_x\\
      iD_x&2J&-J_z&-iD_x\\
      0&-iD_x&iD_x&J_z
      \end{array}
   \right) .
\end{equation}
This matrix is centrosymmetric.
Because the sums and products of CS matrices are again the CS
matrix, the corresponding Gibbs density matrix is CS one.
Therefore,
% instead of numerical calculations \cite{CZ10}
the thermal quantum discord is found here in an exact analytical
form.

\begin{figure}[t]
\begin{center}
\input{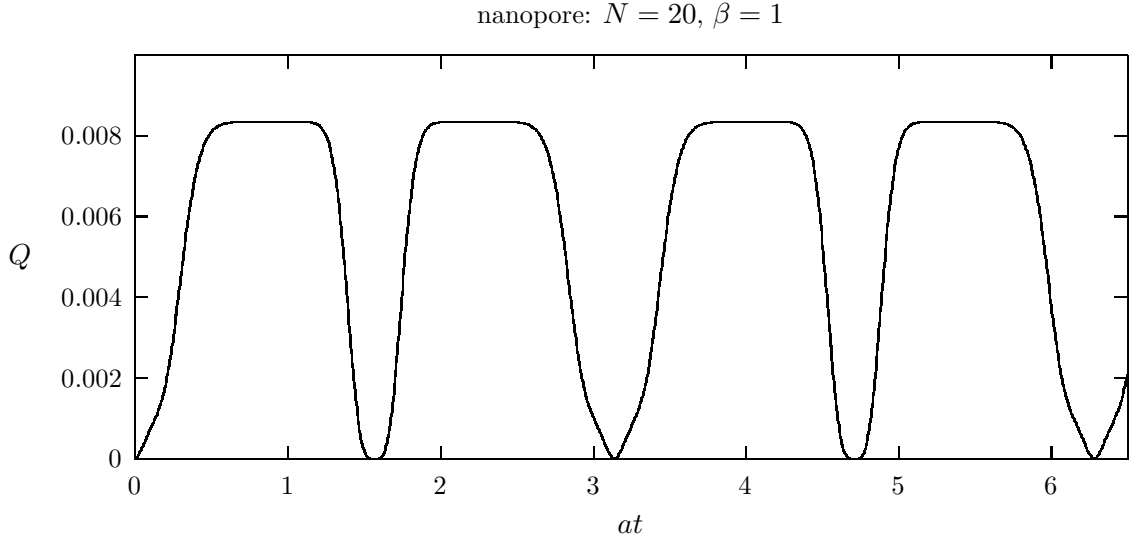}
\caption{Quantum discord in nanopore as a function of dimensionless
time.}
\label{fig:Qt}
\end{center}
\end{figure}

Another example is related to the dynamics in NMR of quantum
correlation (discord) for the pair of nuclear spins in a nanopore
filled with a gas of spin-carrying molecules or atoms
\cite{FKY12}.
The corresponding reduced density matrix is also the CS one:
\begin{equation}
   \label{eq:rho-pore}
   \rho=\left(
      \begin{array}{cccc}
      {1\over4}&{1\over2}p-iu&{1\over2}p-iu&q-r\\
      {1\over2}p+iu&{1\over4}&q+r&{1\over2}p+iu\\
      {1\over2}p+iu&q+r&{1\over4}&{1\over2}p+iu\\
      q-r&{1\over2}p-iu&{1\over2}p-iu&{1\over4}
      \end{array}
   \right),
\end{equation}
where the correlation functions equal 
\begin{eqnarray}
   \label{eq:pqru}
   &&p={1\over2}\tanh{\beta\over2}\cos^{N-1}(at),
   \nonumber\\
   &&q={1\over8}\tanh^2{\beta\over2}[1+\cos^{N-2}(2at)],
   \nonumber\\
   &&r={1\over8}\tanh^2{\beta\over2}[1-\cos^{N-2}(2at)],
   \\
   &&u={1\over4}\tanh{\beta\over2}\cos^{N-2}(at)\sin(at).
   \nonumber
\end{eqnarray}
In these relations, $N$ is the number of particles confined in a nanopore,
$a$ is the normalized coupling constant,
and $\beta$ is the inverse dimensionless temperature.
In the paper \cite{FKY12}, it was succeeded to calculate the quantum
discord only for a particular case $p=u=0$ and $q=r$ when the density
matrix (\ref{eq:rho-pore}) is reduced to the Bell-diagonal form.  
Applying the method developed above we see that after performing the
Hadamard transformation (\ref{eq:CS-X}), the matrix (\ref{eq:rho-pore})
takes the X structure 
\begin{equation}
   \label{eq:rho1-pore}
   \rho^{\,\prime}=\left(
      \begin{array}{cccc}
      {1\over4}+p+q&0&0&-r+2iu\\
      0&{1\over4}-q&r&0\\
      0&r&{1\over4}-q&0\\
      -r-2iu&0&0&{1\over4}-p+q
      \end{array}
   \right),
\end{equation}
or in the Bloch form:
\begin{equation}
   \label{eq:rho1-Bloch}
   \rho^{\,\prime}={1\over4}\,[1 + 2p(\sigma^z_1 + \sigma^z_2)
   + 4r\sigma^y_1\sigma^y_2 + 4q\sigma^z_1\sigma^z_2 
   - u(\sigma^x_1\sigma^y_2 + \sigma^y_1\sigma^x_2)] .
\end{equation}
Using the general formulas of Ref.~4 we can in principle get the discord.
But there is a simpler way.
Indeed, perform an additional local unitary transformation to eliminate
`$xy$' cross-terms in Eq.~(\ref{eq:rho1-Bloch}), that is to reduce
the density matrix (\ref{eq:rho1-pore}) to the {\em real\/} X form.
We are achieving this goal with the transformation
$U\otimes U\rho^{\,\prime}U^+\otimes U^+\equiv\rho^{\,\prime\prime}$,
where $U=\exp(-i\varphi\sigma_z/2)$ and
$\varphi=-{1\over2}\arctan(2u/r)$.
After this transformation, the density matrix (\ref{eq:rho1-pore})
takes the form
\begin{equation}
   \label{eq:rho11-pore}
   \rho^{\,\prime\prime}=\left(
      \begin{array}{cccc}
      {1\over4}+p+q&0&0&2u\sin2\varphi-r\cos2\varphi\\
      0&{1\over4}-q&r&0\\
      0&r&{1\over4}-q&0\\
      2u\sin2\varphi-r\cos2\varphi&0&0&{1\over4}-p+q
      \end{array}
   \right),
\end{equation}
that is
\begin{eqnarray}
   \label{eq:rho11-Bloch}
   \rho^{\,\prime\prime}&=&{1\over4}\,[1 + 2p(\sigma^z_1 + \sigma^z_2)
   + 4(r\sin^2\varphi+u\sin2\varphi)\sigma^x_1\sigma^x_2 
   + 4(r\cos^2\varphi-u\sin2\varphi)\sigma^y_1\sigma^y_2 
   \nonumber\\
   &+& 4q\sigma^z_1\sigma^z_2] . 
\end{eqnarray}
Using now the formulas for calculating the quantum discord for the
real X density matrices \cite{F10} we finally obtain
\begin{equation}
   \label{eq:Q-Q1Q2}
   Q = \min\{Q_1, Q_2\} , 
\end{equation}
where
\begin{eqnarray}
   \label{eq:Q1}
   Q_1&=&S_r-S
   -({1\over4}+p+q)\log_2\frac{{1\over4}+p+q}{{1\over2}+p}
   -({1\over4}-q)\log_2\frac{{1\over4}-q}{{1\over2}+p}
   \nonumber\\
   &-&({1\over4}-p+q)\log_2\frac{{1\over4}-p+q}{{1\over2}-p}
   -({1\over4}-q)\log_2\frac{{1\over4}-q}{{1\over2}-p} ,
\end{eqnarray}
\begin{equation}
   \label{eq:Q2}
   Q_2 = S_r-S-D_1\log_2D_1-D_2\log_2D_2 , 
\end{equation}
and
\begin{equation}
   \label{eq:D12}
   D_{1,2} = {1\over2}\,\lbrack\!\lbrack1\pm2[p^2+(|r|
   +|2u\sin2\varphi-r\cos2\varphi|)^2]^{1/2}\rbrack\!\rbrack . 
\end{equation}
In Eqs.~(\ref{eq:Q1}) and (\ref{eq:Q2}), $S$ and $S_r$ are the
entropies of the full and reduced density matrices, respectively:
\begin{equation}
   \label{eq:S}
   S = -\sum_{j=1}^4\lambda_j\log_2\lambda_j ,
\end{equation}
where the eigenvalues $\lambda_j$ of the density matrix under
consideration are given as
\begin{equation}
   \label{eq:l1-l4}
   \lambda_{1,2} = {1\over4}+q\pm[p^2
   +(2u\sin2\varphi-r\cos2\varphi)^2]^{1/2} , \qquad 
   \lambda_{3,4} = {1\over4}-q\pm|r|,
\end{equation}
and
\begin{equation}
   \label{eq:Sr}
   S_r = -({1\over2}+p)\log_2({1\over2}+p)
         -({1\over2}-p)\log_2({1\over2}-p) .
\end{equation}
Typical time dependence for the pairwise quantum correlation
of spin-carrying particles in a nanopore is shown in Fig.~\ref{fig:Qt}.
The correlation oscillates between zero and the saturation value which
is $Q$ when $p=u=0$ and $r=q={1\over8}\tanh^2(\beta/2)$ [see Eq.~(26)
in Ref.~7].
%......................................................................
%                          FIGURE 1
%......................................................................

Our third example concerns with the pseudopure (PP) states
\begin{equation}
   \label{eq:rho-PP}
   \rho_{PP}=\alpha|\psi\rangle\langle\psi| + \frac{1-\alpha}{4}I ,
\end{equation}
where $|\psi\rangle$ is an arbitrary two-qubit pure state, $I$ is the
identity operator, and the probability $\alpha\in[0,1]$.
The states like (\ref{eq:rho-PP}) are studied as a possible resource
for NMR quantum computing (see, e.~g., \cite{MAC12} and references
therein).
It is easy to check that if
\begin{equation}
   \label{eq:psi}
   |\psi\rangle=a(|00\rangle + |11\rangle)
   + b(|01\rangle + |10\rangle)
\end{equation}
($|a|^2+|b|^2=1/2$ and $\{|00\rangle, |01\rangle, |10\rangle,    
|11\rangle\}$ is the computational basis) the state
$\rho_{PP}$ will be CS one and we can calculate its quantum discord
using the above scheme.
 
{\em Conclusions.\/} We have established the relation between
the CS and X matrices via the universal local orthogonal transformation.
This allows to find the discord of any CS states using available formulas
for the discord of X states.
CS quantum states appear in different important physical problems,
three of which have been discussed. 

\vspace{3mm}
This work was supported by the RFBR (project No.~13-03-00017).
%This research was
%% partially
% supported in part by the program No.~8 of the
%Presidium of RAS.

%======================================================================

%======================================================================

\end{document}